\documentclass[a4paper,11pt]{article}
\usepackage{pos}
\usepackage{subcaption}
\usepackage{bookmark}
\usepackage{float}
\usepackage[percent]{overpic}
\usepackage[nice]{nicefrac}
\usepackage[
  separate-uncertainty=true,
]{siunitx}
\title{Computing sky maps using the open-source package Gammapy and MAGIC data in a standardized format}
\ShortTitle{Computing sky maps using Gammapy and MAGIC data}

\author*[a]{Simone Mender}
\author[a]{Lena Linhoff}
\author[b]{Tarek Hassan}
\author[c]{Cosimo Nigro}
\author[a]{Dominik Elsässer}

\affiliation[a]{TU Dortmund University,\\
  Otto-Hahn-Straße 4a, 44227 Dortmund, Germany}

\affiliation[b]{Centro de Investigaciones Energéticas, Medioambientales y Tecnológicas,\\
E-28040 Madrid, Spain}

\affiliation[c]{Institut de Física d’Altes Energies (IFAE), The Barcelona Institute of Science and Technology,\\
Campus UAB, Bellaterra, 08193 Barcelona, Spain}

\emailAdd{simone.mender@tu-dortmund.de}
\emailAdd{lena.linhoff@tu-dortmund.de}
\emailAdd{tarek.hassan@ciemat.es}
\emailAdd{dominik.elsaesser@tu-dortmund.de}
\emailAdd{cosimo.nigro@ifae.es}

\abstract{
  The open-source Python package Gammapy, developed for the high-level analysis of gamma-ray data,
  requires gamma-like event lists combined with the corresponding instrument response functions.
  For a morphological analysis, these data have to include a background acceptance model.
  Here we report an approach to generate such a model for the MAGIC telescope data,
  accounting for the azimuth and zenith dependencies of the MAGIC background acceptance.
  We validate this method using observations of the Crab Nebula with different offsets from the pointing position.
}

\FullConference{%
  7th Heidelberg International Symposium on High-Energy Gamma-Ray Astronomy (Gamma2022)\\
  4-8 July 2022\\
  Barcelona, Spain\\}

\begin{document}
\maketitle

\section{Introduction}
So far, the analysis of MAGIC data is traditionally performed with the proprietary software \texttt{MARS}\citep{mars}.
For the future generation of very-high-energy gamma-ray instruments, 
the gamma-ray community is developing standardized data formats \textit{Data Formats for Gamma-ray Astronomy} (GADF)\citep{universe7100374},
and Gammapy\citep{gammapy:2017}, an open-source tool, compatible with this data format.

In this contribution, we present an approach to create a background acceptance model
for MAGIC data, which accounts for the zenith distance dependence of the MAGIC background.
The background models created in this way are then used to create skymaps with Gammapy
and MAGIC data for the first time.
For validation, Crab Nebula datasets with different offsets to the pointing position are analyzed.

\section{Characterisation of the MAGIC gamma-like acceptance}
\begin{figure}[h]
  \begin{subfigure}[b]{0.4\textwidth}
    \centering
    \includegraphics[width=\textwidth]{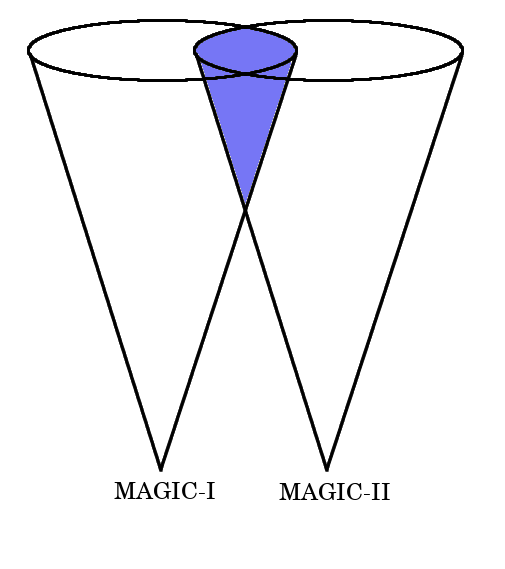}
 \end{subfigure}
  \begin{subfigure}[b]{0.6\textwidth}
   \centering
   \begin{overpic}[width=\textwidth]{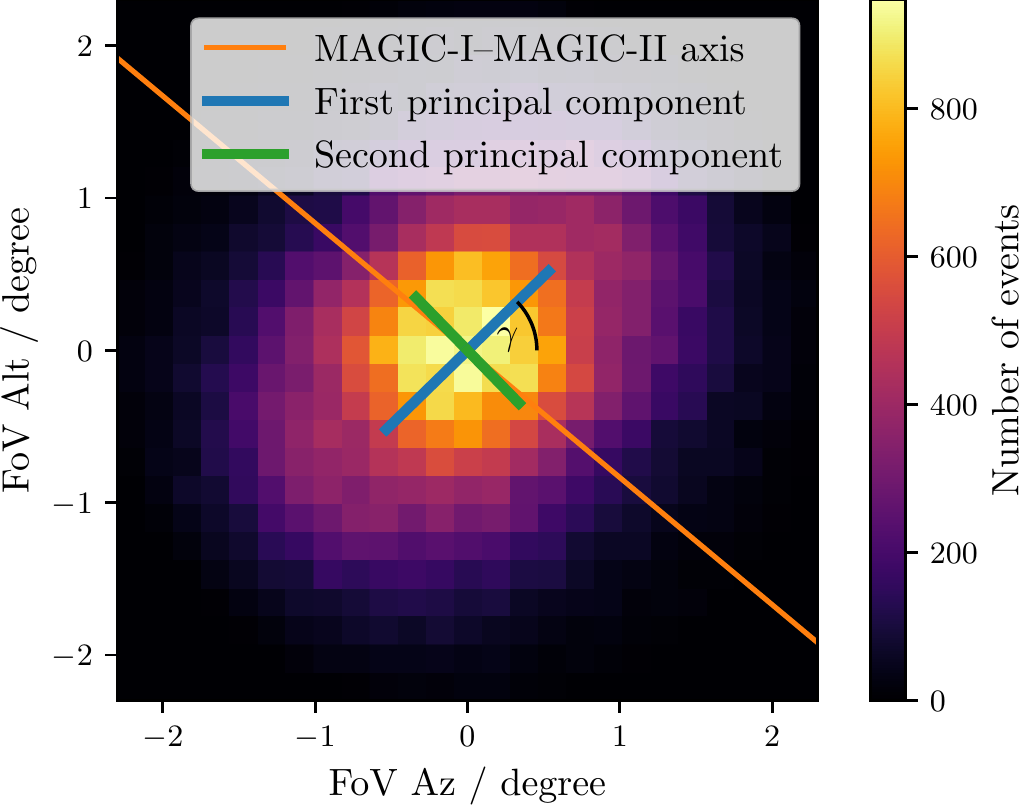}
    \put (15,12) {\color{gray} Preliminary}
   \end{overpic}
\end{subfigure}
\caption{
Left: 
Schematic illustration of the viewing cones of MAGIC-I and MAGIC-II.
The overlapping part is coloured in blue. 
Right: 
The plot shows an exemplary azimuth bin from \SI{69}{\degree} to \SI{99}{\degree}.
A principle component analysis is applied to the events in each azimuth bin.
The angle between the horizontal and the first principal component (blue line) is called rotation angle $\gamma$.}
\label{fig:rotation-angle}
\end{figure}
In a previous study\citep{inproceedings}, it was found that the shape of the MAGIC acceptance
is rotating dependent on the azimuth angle of the pointing position of the observation.

Here, we present an explanation for the azimuth-dependent rotation and compare this theory with the data.
In the left part of Figure \ref{fig:rotation-angle}, the viewing cones of MAGIC-I and MAGIC-II are 
illustrated.
An air shower has to pass both viewing cones to be triggered.
As the non-symmetrical overlapping part of both viewing cones has its broadest part 
perpendicular to the connection line between both telescopes,
it is predicted that events from that direction will be triggered preferentially.
Projected into the sky, the overlapping part of the viewing cones is rotating 
dependent on the azimuth angle of the observation.
In this analysis, the azimuth angle $Az$ is counted from the Geographic North ($Az=\SI{0}{\degree}$) 
to East ($Az=\SI{90}{\degree}$).
The angle $\alpha$ between the North-South axis and the MAGIC-I{\textendash}MAGIC-II axis is calculated as
\begin{equation}
  \alpha = \arctan((\overrightarrow{M_1} - \overrightarrow{M_2})_y / (\overrightarrow{M_1} - \overrightarrow{M_2})_x) \approx \SI{34.23}{\degree} 
\end{equation}
with the positions $\overrightarrow{M_1}$ and $\overrightarrow{M_2}$ of the two telescopes:
\begin{equation}
  \overrightarrow{M_1} = \left(41.054, -79.275\right)^\top\,\mathrm{m} \mathrm{~~and~~} 
  \overrightarrow{M_2} = \left(-29.456, -31.295\right)^\top\,\mathrm{m}.
\end{equation}
In this definition, the $y$-axis points towards North, the $x$-axis points towards East
and the coordinate origin is the dish-area-weighted center of MAGIC-I, MAGIC-II and the LST-1.
As the overlapping part of the two viewing cones has its broadest part orthogonal to the MAGIC-I{\textendash}MAGIC-II axis,
the theoretical expected function of the rotation angle is:
\begin{equation}
  \label{eq:theory}
  \gamma_\mathrm{theory} (Az) = Az - \alpha \approx Az - \SI{34.23}{\degree}.
\end{equation}

\begin{figure}[H]
   \centering
   \begin{overpic}[width=0.65\textwidth]{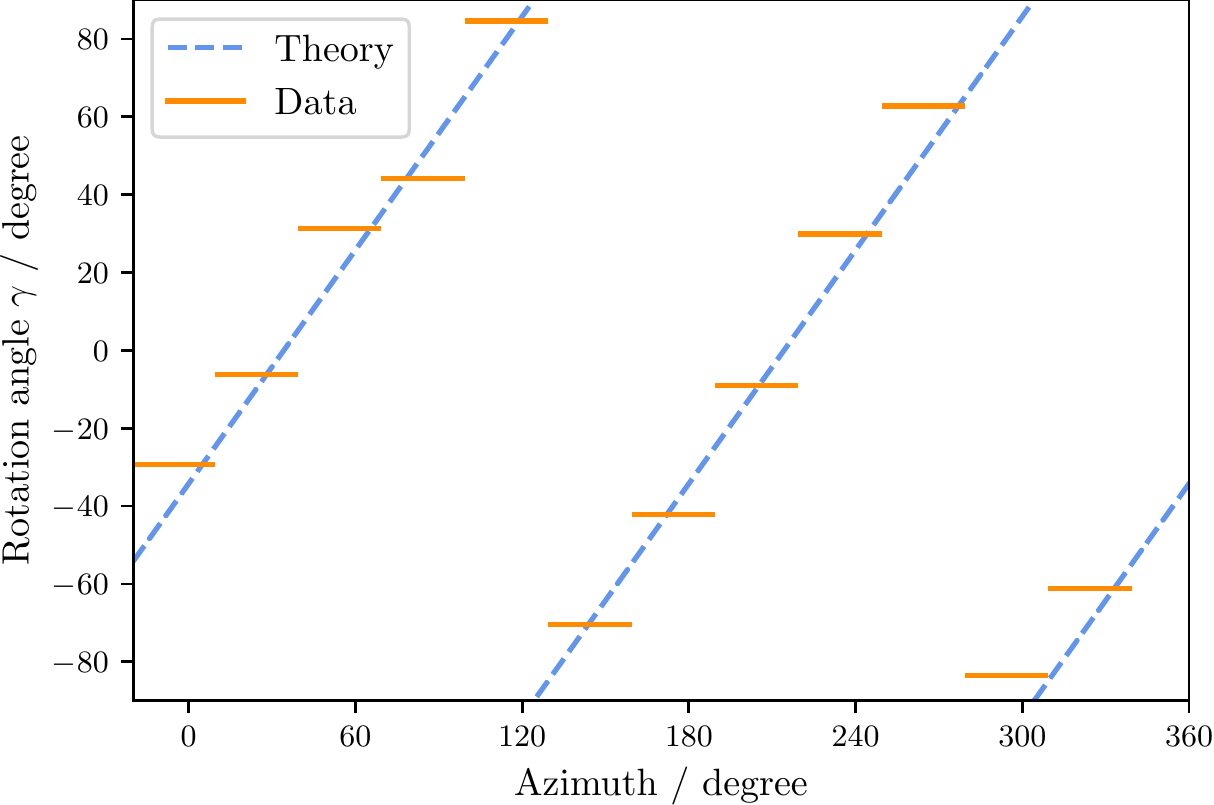}
    \put (15,10) {\color{gray} Preliminary}
   \end{overpic}
\caption{
Rotation angle $\gamma$ dependency on the azimuth angle of an observation.
}
\label{fig:gamma-vs-az}
\end{figure}

To compare this assumption with observations, a large DL3\footnote[1]{Data Level 3 (DL3) data format according to GADF: \url{https://gamma-astro-data-formats.readthedocs.io/}} dataset of off observations is analyzed in twelve azimuth bins.
The estimated energy of those gamma-like events ranges from \SI{0.05}{\TeV} to \SI{10}{\TeV}.
In each azimuth bin, a principal component analysis is performed on the events in field-of-view (FoV) coordinates.
The first principal component is indicating the direction of the greatest variance in the data and is therefore expected to 
be perpendicular to the connection line of the two telescopes.
In the right part of Figure \ref{fig:rotation-angle}, this procedure is visualized in
an exemplary azimuth bin from \SI{69}{\degree} to \SI{99}{\degree}.
In Figure \ref{fig:gamma-vs-az}, the comparison between the measured rotation angles from the data
and the theory curve \eqref{eq:theory} shows a good agreement.

\begin{figure}[H]
  \begin{subfigure}[b]{0.5\textwidth}
   \centering
   \begin{overpic}[width=0.9\textwidth]{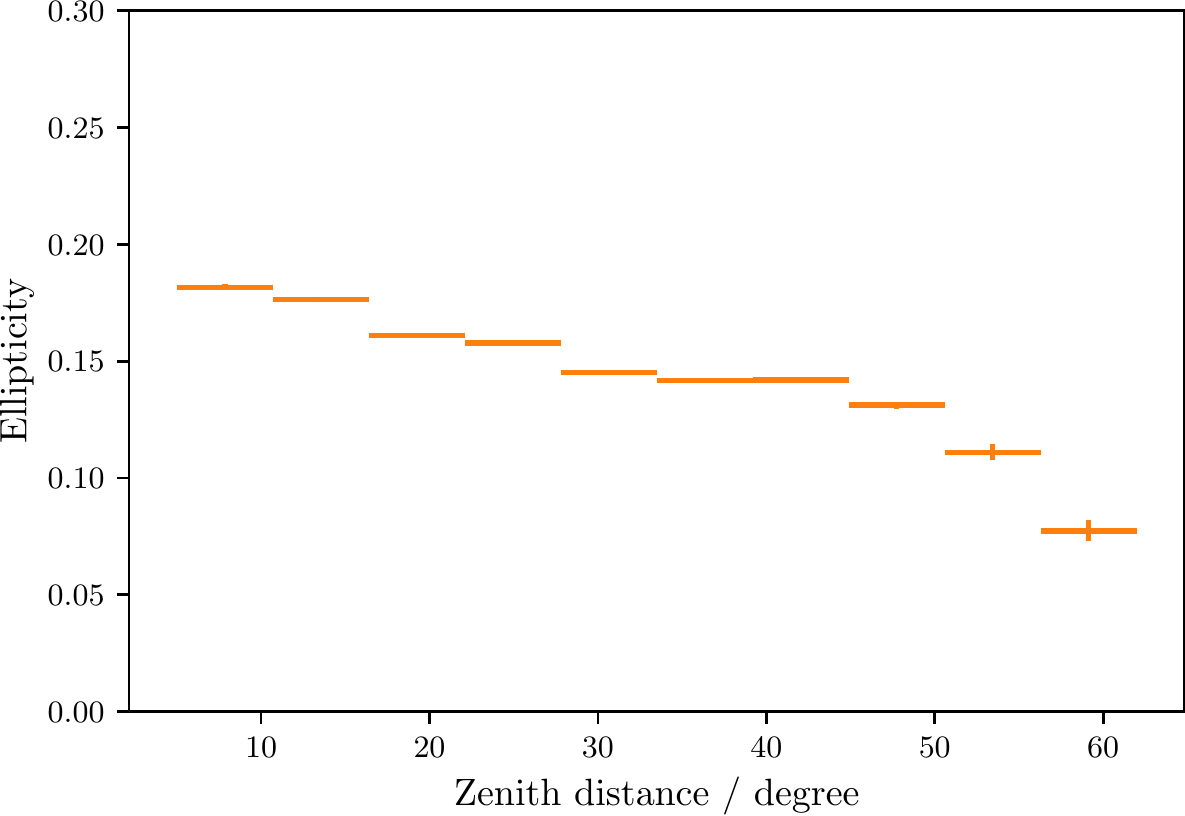}
    \put (15,10) {\color{gray} Preliminary}
   \end{overpic}
\end{subfigure}
       \begin{subfigure}[b]{0.5\textwidth}
   \centering
   \begin{overpic}[width=0.9\textwidth]{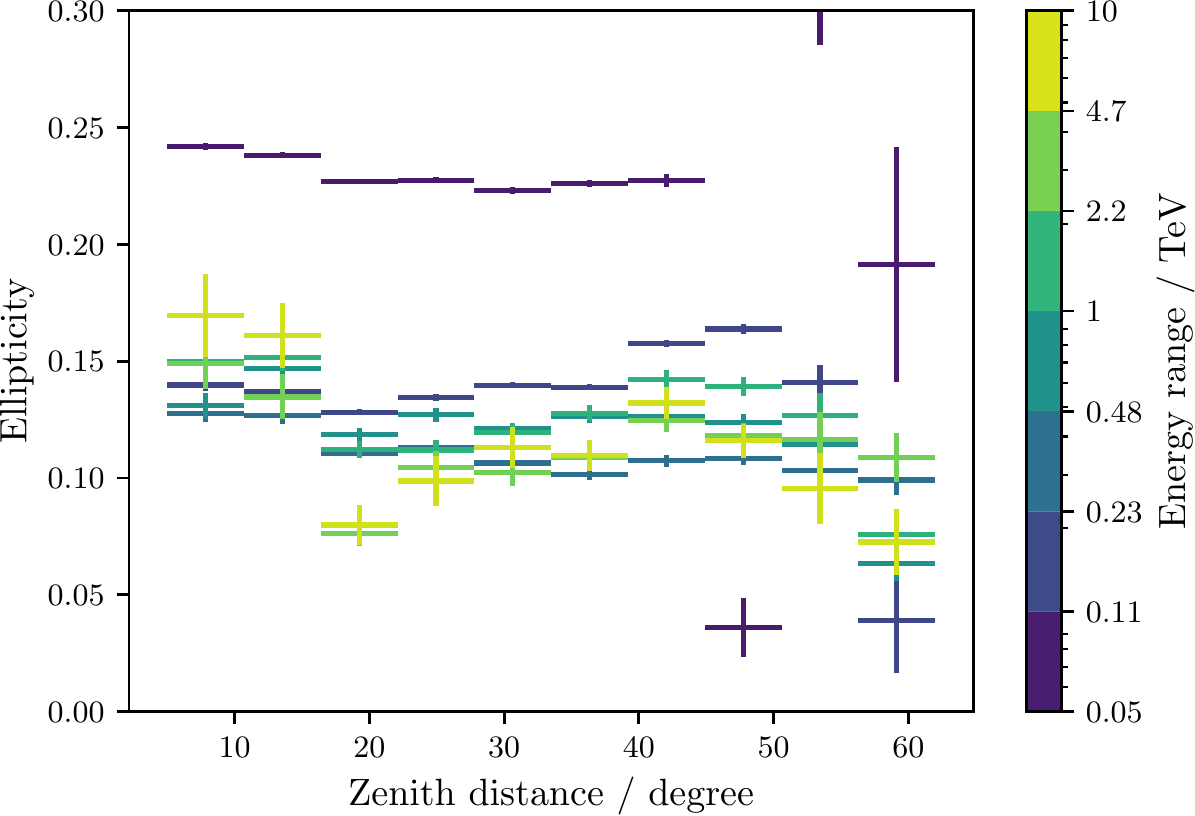}
    \put (15,10) {\color{gray} Preliminary}
   \end{overpic}
\end{subfigure}
\caption{
Ellipticity of the gamma-like acceptance as a function of the zenith distance
in one energy bin (left) and splitted in seven energy bins (right) from \SI{50}{\MeV} to \SI{10}{\TeV}.
The errorbars in these plots are calculated using a bootstrap approach.
}
\label{fig:ellipticity}
\end{figure}

To study the zenith dependency, the first-order influence of the azimuth can be removed by a correction.
  The FoV coordinates of the events are de-rotated by the theoretical rotation angle $\gamma_\text{theory}(Az_\text{Pointing})$ 
  of the corresponding pointing position: 
  \begin{equation}
    \label{eq:derotation}
    \left(\begin{array}{cc}
   \text{lon}_\text{rotated} \\
   \text{lat}_\text{rotated} \\
  \end{array} \right) = 
          \left(\begin{array}{cc}
    \cos(\gamma_\text{theory})& -\sin(\gamma_\text{theory})\\
   \sin(\gamma_\text{theory}) & \cos(\gamma_\text{theory}) \\
  \end{array} \right) \cdot
      \left(\begin{array}{cc}
   \text{lon} \\
   \text{lat} \\
  \end{array} \right)
  \end{equation}
On the de-rotated event coordinates, again a principal component analysis is performed in different zenith bins.
With the resulting semi-major axis of the covariance ellipse $a$ and
the semi-minor axis of the covariance ellipse $b$ the ellipticity 
\begin{equation}
  \epsilon = \frac{\sqrt{a}-\sqrt{b}}{\sqrt{a}}
\end{equation}
can be calculated.
As visible in the left part of Figure \ref{fig:ellipticity},
the ellipticity is decreasing with higher zenith distances, which confirms the results of \citep{inproceedings}.
Furthermore, we investigated the ellipticity in different energy bins.
At lower energies, the ellipticity of the acceptance is higher
as shown in the right part of Figure \ref{fig:ellipticity}.
This indicates that the effect of a higher ellipticity at a lower zenith distance may 
not be an effect of the zenith distance, but of the energy of the events,
because at higher zenith distances the data contains a higher percentage of high-energy events.
In a certain energy bin -- especially at low energies -- the ellipticity is constant over a large zenith range.
At higher energies, there is less statistics, but no clear trend is visible.

\section{Creation of skymaps with Gammapy}
\label{sec:skymaps}
\begin{figure}[H]
  \begin{subfigure}[b]{0.49\textwidth}
   \centering
   \begin{overpic}[width=\textwidth]{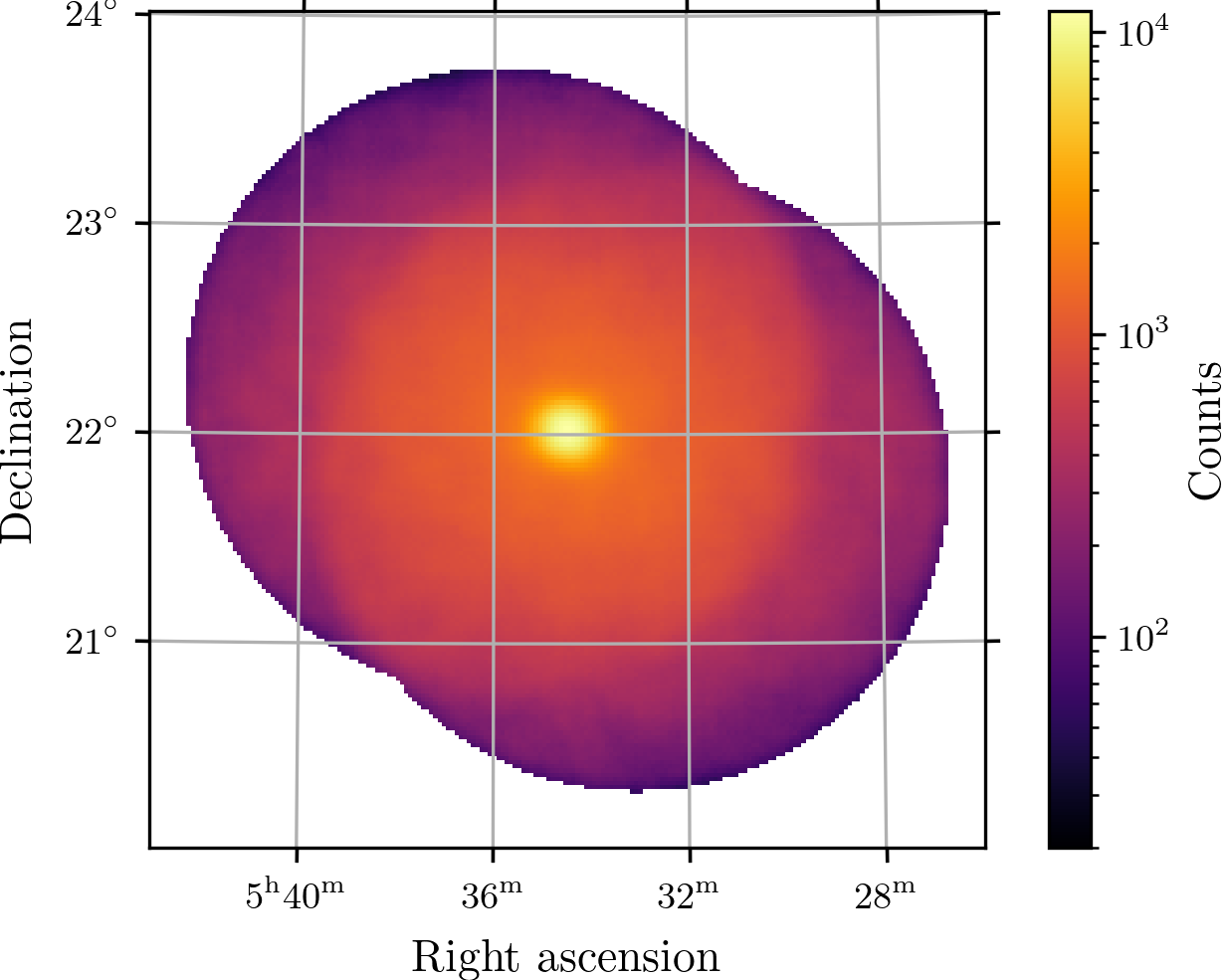}
    \put (15,12) {\color{gray} Preliminary}
   \end{overpic}
\end{subfigure}
       \begin{subfigure}[b]{0.49\textwidth}
   \centering
   \begin{overpic}[width=\textwidth]{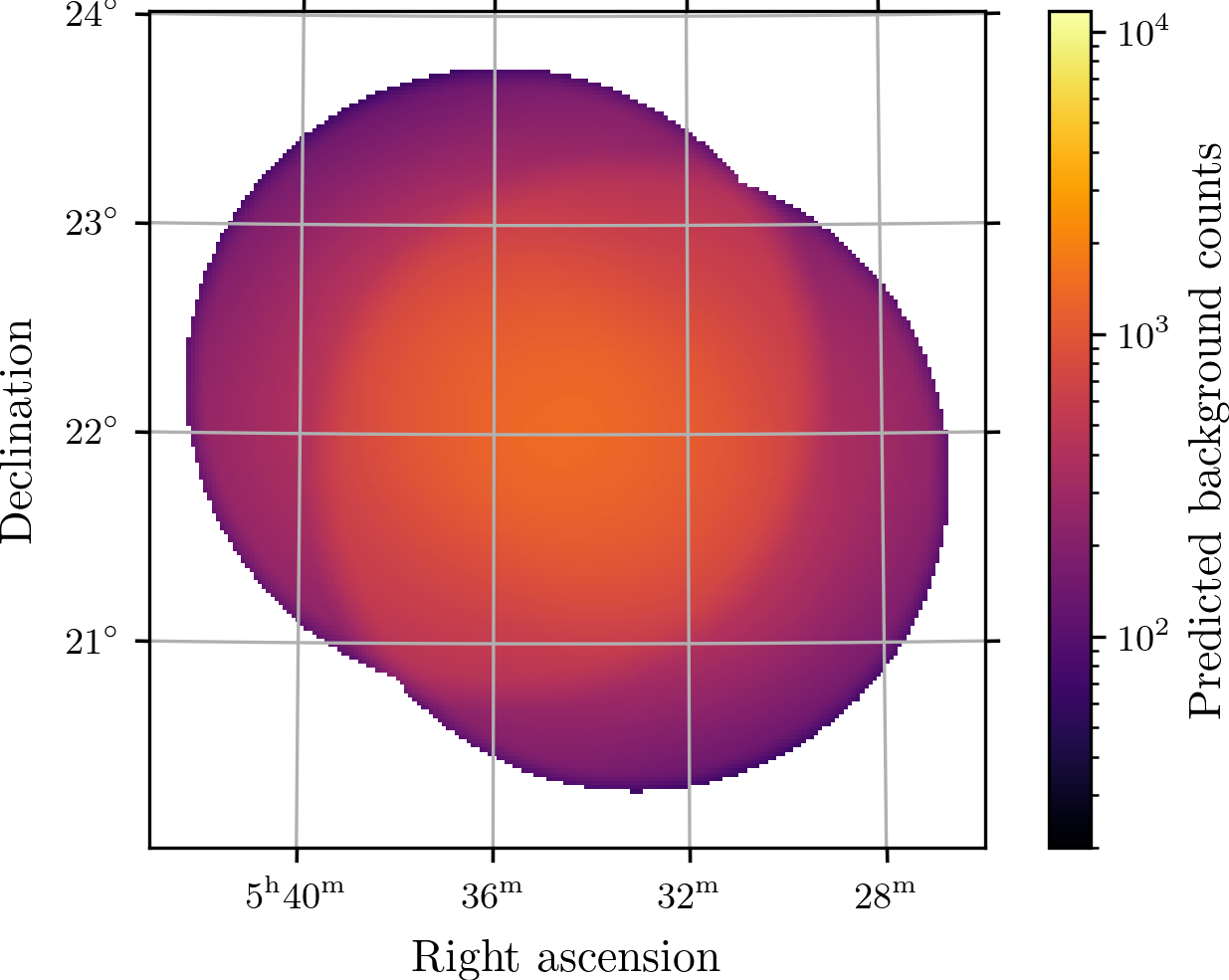}
    \put (15,12) {\color{gray} Preliminary}
   \end{overpic}
\end{subfigure}
\caption{Counts map (left) and predicted background counts map (right) for a stacked Crab Nebula dataset.}
\label{fig:counts-map}
\end{figure}

\begin{figure}[H]
  \begin{subfigure}[b]{0.49\textwidth}
\centering
\begin{overpic}[width=\textwidth]{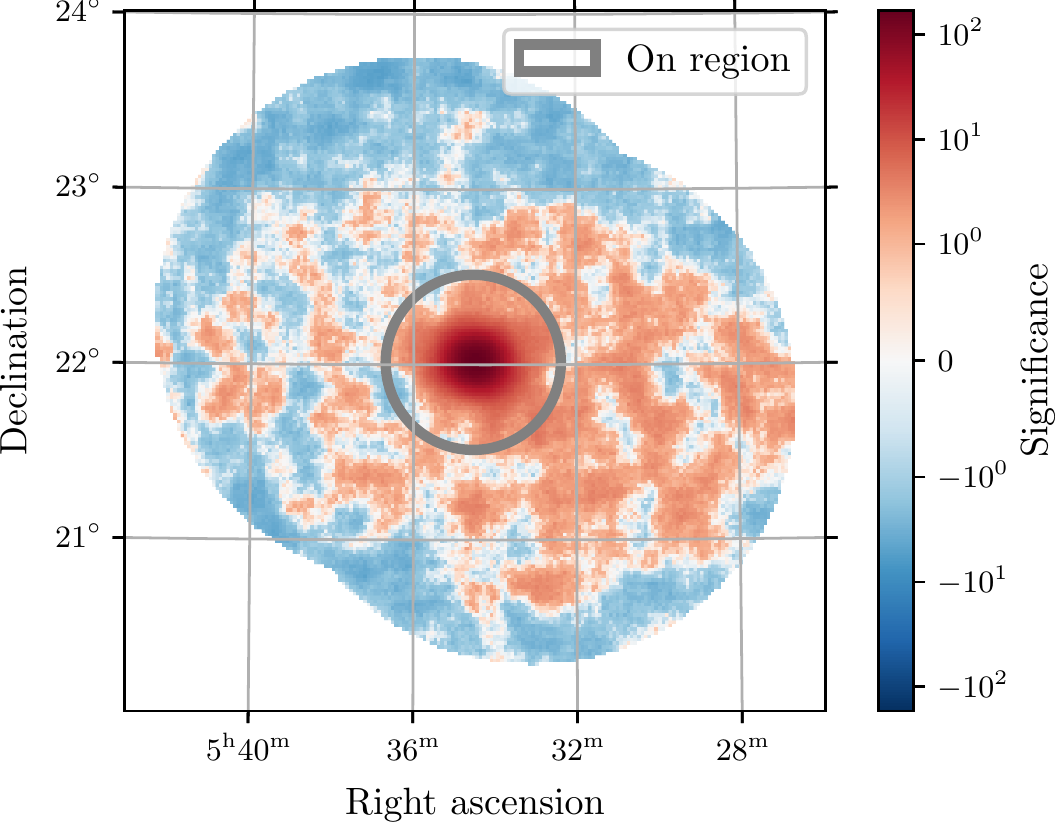}
  \put (15,12) {\color{gray} Preliminary}
 \end{overpic}
\end{subfigure}
\begin{subfigure}[b]{0.49\textwidth}
\centering
\begin{overpic}[width=\textwidth]{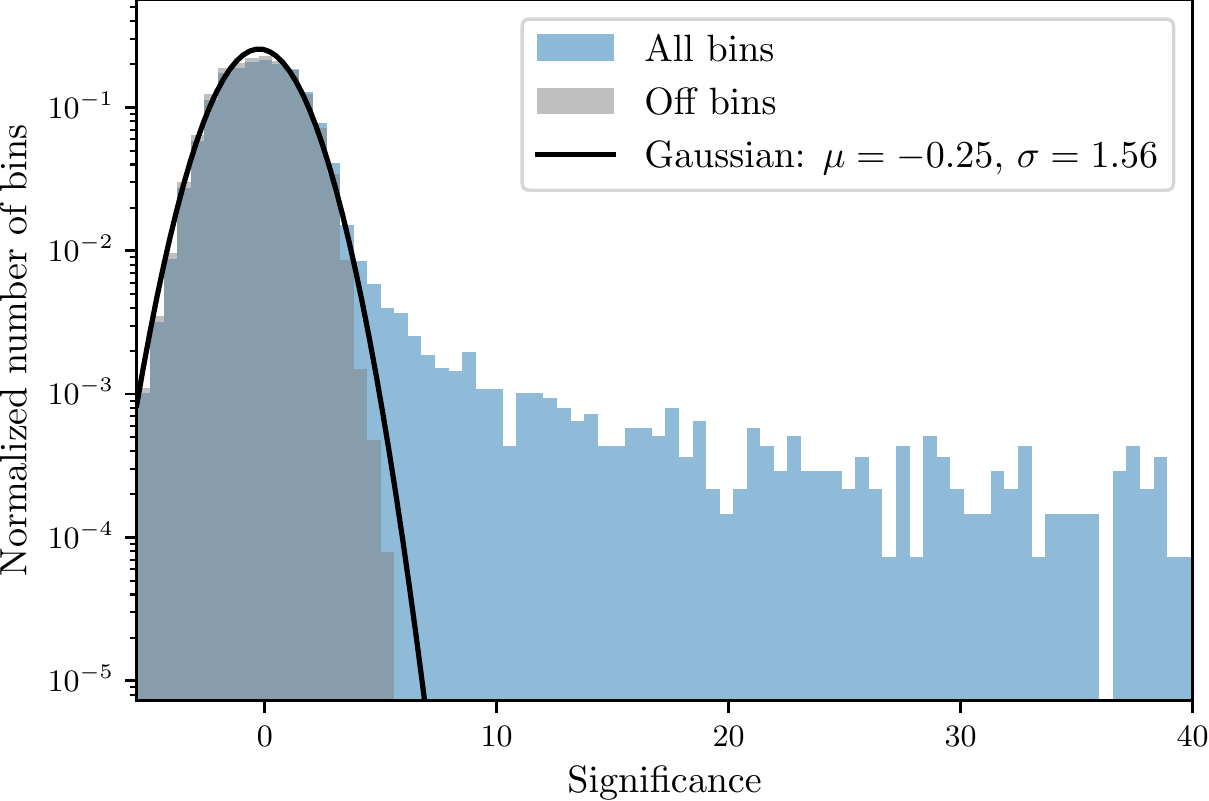}
  \put (50,10) {\color{gray} Preliminary}
 \end{overpic}
\end{subfigure}
\caption{Excess significance map (left) for a stacked Crab Nebula dataset and corresponding significance value histogram (right).}
\label{fig:significance-map}
\end{figure}

To perform a morphological analysis with Gammapy\citep{gammapy:2017}, the data has to contain an acceptance model.
With the knowledge about the azimuth dependence,
we create acceptance models from an off dataset with the following procedure:
\begin{enumerate}
  \item De-rotate the position of all events from off observations around the predicted theory rotation angle \eqref{eq:theory}
  of the corresponding pointing position.
\item  Rotate the position of all events around the theory rotation angle \eqref{eq:theory} of the pointing position from the observation for
which the acceptance should be calculated.
\item  Histogram events and calculate background rate in $\mathrm{s}^{-1}\,\mathrm{MeV}^{-1}\,\mathrm{sr}^{-1}$.
\end{enumerate}
The zenith dependency is implicitly taken into account by performing this procedure in different energy bins.
Finally, the data is stored according to the GADF as \texttt{3DBackground} model.
Gammapy offers the opportunity to normalize the background acceptance model
to a single observation with the \texttt{FoVBackgroundMaker} (fit method and scale method) and the \texttt{RingBackgroundMaker}.
Multiple observations can be stacked to a single \texttt{MapDataset},
from which the \texttt{ExcessMapEstimator} can produce skymaps.
Here, we present the results of the \texttt{FoVBackgroundMaker} (fit method) applied to a Crab Nebula dataset.
The data contains observations measured within a zenith range of \SI{5}{\degree} to \SI{35}{\degree} with a standard wobble offset of \SI{0.4}{\degree}.
In Figure \ref{fig:counts-map}, the resulting counts map and the predicted background counts maps are shown.
The significance map and the corresponding significance value histrogram are shown in Figure \ref{fig:significance-map}.

\section{Validation of the spectrum estimation using multiple Crab Nebula datasets}
\begin{figure}[H]
  \begin{subfigure}[b]{0.5\textwidth}
\centering
\begin{overpic}[width=\textwidth]{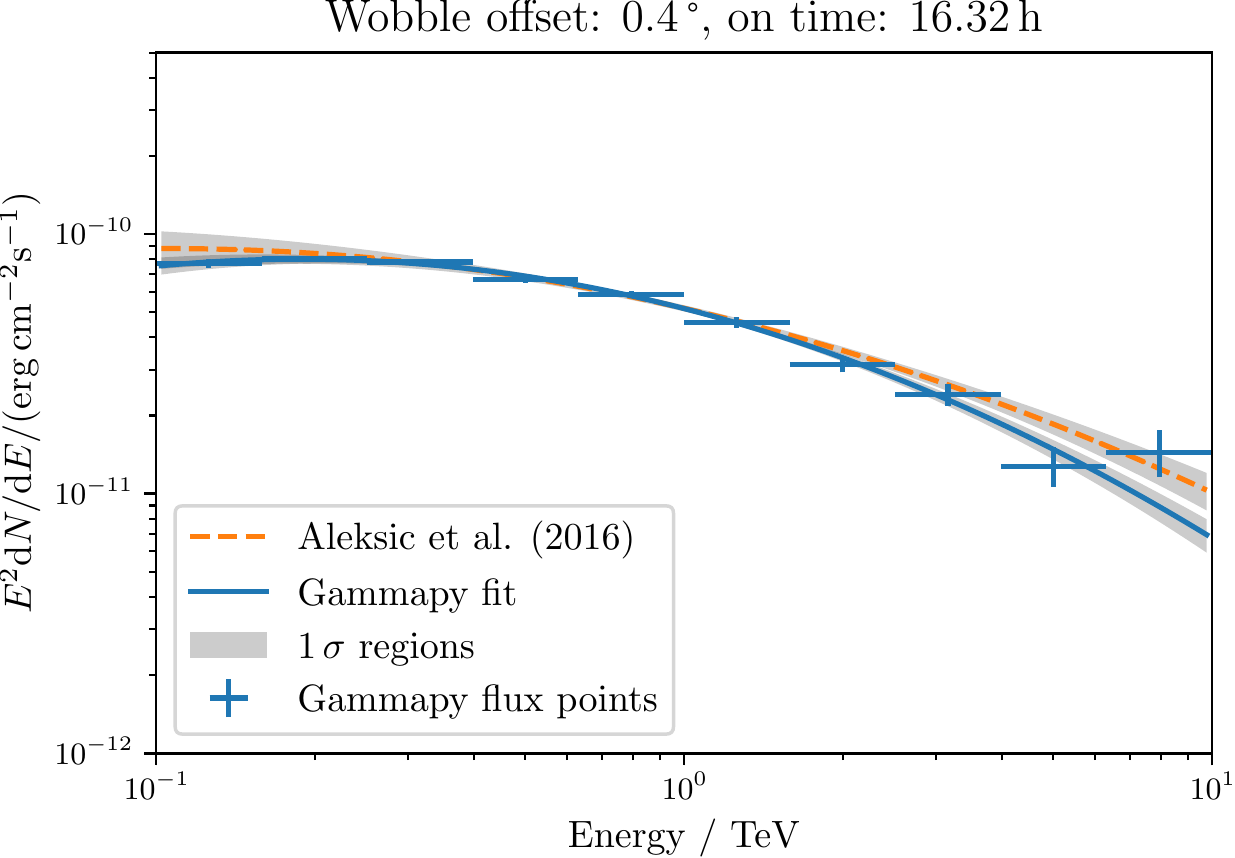}
  \put (70,10) {\color{gray} Preliminary}
 \end{overpic}
\end{subfigure}
\begin{subfigure}[b]{0.5\textwidth}
\centering
   \begin{overpic}[width=\textwidth]{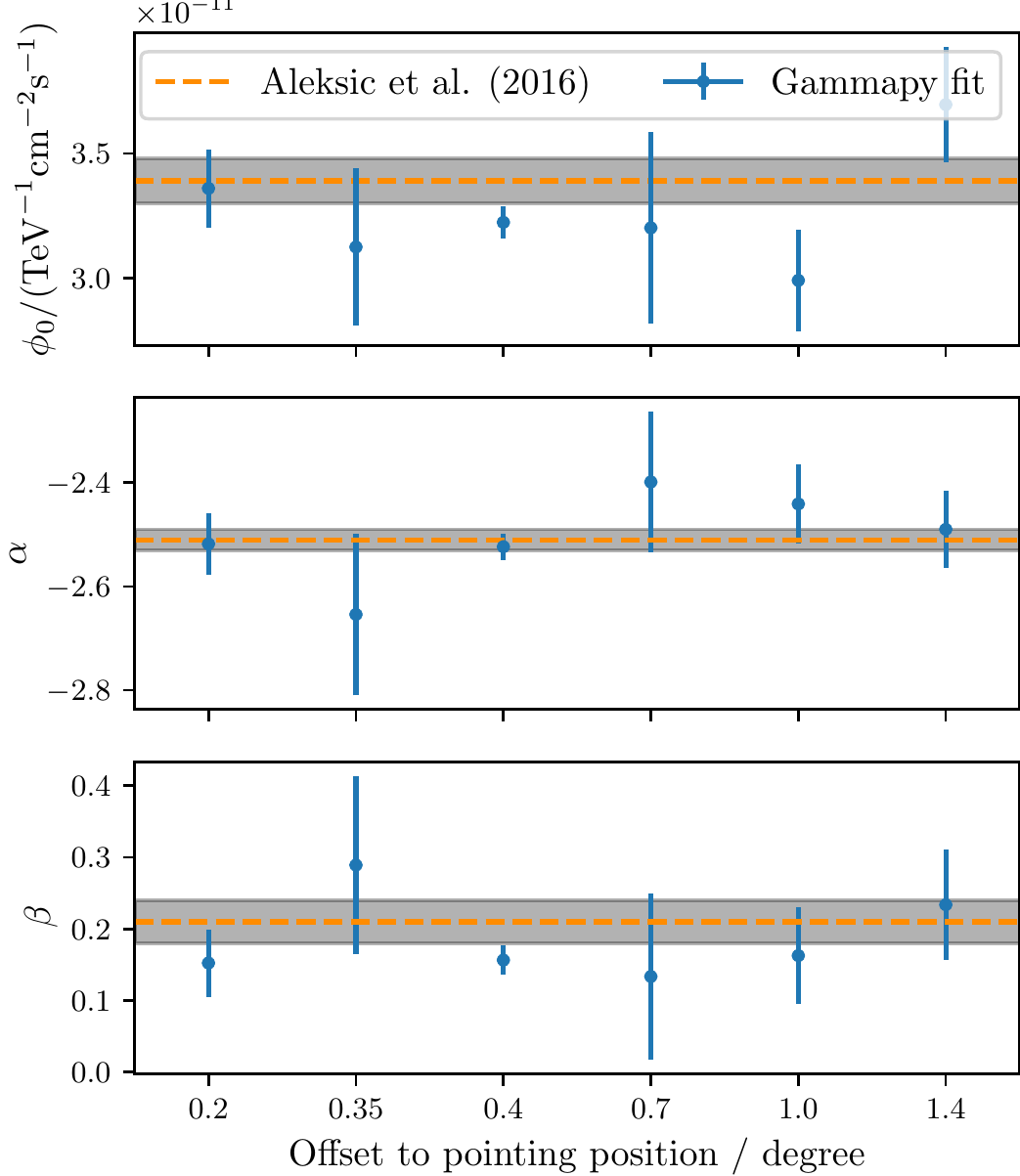}
    \put (15,10) {\color{gray} Preliminary}
   \end{overpic}
\end{subfigure}
\caption{
Left: Spectral energy distribution from a Crab Nebula observation with an offset of \SI{0.4}{\degree}.
Right: Estimated parameters of the log-parabola model \eqref{eq:log-parabola} for different wobble offsets.
}
\label{fig:spectra}
\end{figure}

Based on the analysis in section \ref{sec:skymaps}, 
we estimated a Crab Nebula spectrum assuming a log-parabola model 
\begin{equation}
  \label{eq:log-parabola}
  \phi(E)=\phi_0 \left(\frac{E}{\SI{1}{\TeV}}\right)^{-\alpha-\beta\log\left(\nicefrac{E}{\SI{1}{\TeV}}\right)}
\end{equation}
with the forward-folding likelihood analysis implemented in Gammapy.
To validate the methods in the complete field of view,
we analyzed multiple Crab Nebula datasets \citep{ALEKSIC201676} with different offsets to the pointing position.
The data is taken under a zenith angle from \SI{5}{\degree} to \SI{35}{\degree} and
the wobble offset of the observations varies from \SI{0.2}{\degree} to \SI{1.4}{\degree}.
As represented in Figure \ref{fig:spectra}, the resulting log-parabola parameters show good agreement
with the MAGIC reference curve \cite{ALEKSIC201676}. 

\footnotesize
\section*{Ackknowledgements}
This work has been supported by the DFG, Collaborative Research Center SFB 876 under the project C3 
(\url{https://sfb876.tu-dortmund.de}) and the SFB 1491 (\url{https://www.sfb1491.ruhr-uni-bochum.de}).

\bibliographystyle{JHEP}
\bibliography{references.bib}

\providecommand{\href}[2]{#2}\begingroup\raggedright\begin{thebibliography}{1}

\bibitem{mars}
R.~Zanin, \emph{Mars, the magic analysis and reconstruction software},  in
  \emph{Proceedings, 33rd International Cosmic Ray Conference (ICRC2013): Rio
  de Janeiro, Brazil, July 2-9, 2013}, p.~0773, 2013,
  \href{http://inspirehep.net/record/1412925/files/icrc2013-0773.pdf}{http://inspirehep.net/record/1412925/files/icrc2013-0773.pdf}.

\bibitem{universe7100374}
C.~Nigro, T.~Hassan and L.~Olivera-Nieto, \emph{Evolution of data formats in
  very-high-energy gamma-ray astronomy},
  \href{https://doi.org/10.3390/universe7100374}{\emph{Universe} {\bfseries 7}
  (2021) }.

\bibitem{gammapy:2017}
C.~{Deil}, R.~{Zanin}, J.~{Lefaucheur}, C.~{Boisson}, B.~{Khelifi},
  R.~{Terrier} et~al., \emph{{Gammapy - A prototype for the CTA science
  tools}},  in \emph{35th International Cosmic Ray Conference (ICRC2017)},
  vol.~301 of \emph{International Cosmic Ray Conference}, p.~766, Jan., 2017
  [\href{https://arxiv.org/abs/1709.01751}{{\ttfamily 1709.01751}}].

\bibitem{inproceedings}
E.~Prandini, G.~Pedaletti, P.~Vela, E.~Wilhelmi, P.~Colin, C.~Fruck et~al.,
  \emph{Study of hadron and gamma-ray acceptance of the magic telescopes:
  towards an improved background estimation},  p.~721, 08, 2016,
  \href{https://doi.org/10.22323/1.236.0721}{DOI}.

\bibitem{ALEKSIC201676}
J.~Aleksic, S.~Ansoldi, L.~Antonelli, P.~Antoranz, A.~Babic, P.~Bangale et~al.,
  \emph{The major upgrade of the magic telescopes, part ii: A performance study
  using observations of the crab nebula},
  \href{https://doi.org/https://doi.org/10.1016/j.astropartphys.2015.02.005}{\emph{Astroparticle
  Physics} {\bfseries 72} (2016) 76}.

\end{thebibliography}\endgroup

\end{document}